\documentclass[oldversion, letter]
{aa}

\usepackage{amsmath}
\usepackage{natbib}
\usepackage[english]{babel}
\usepackage{epsfig}
\usepackage{txfonts}

\newcommand{\beq}{\begin{equation}}   %

\newcommand{\eeq}{\end{equation}}   %

\newcommand{\beqa}{\begin{eqnarray}}   %

\newcommand{\eeqa}{\end{eqnarray}}   %

\newcommand{\beal}{\begin{align}}

\newcommand{\enal}{\end{align}}

\newcommand{\bspl}{\begin{split}}

\newcommand{\espl}{\end{split}}

\newcommand{\bsub}{\begin{subequations}}

\newcommand{\esub}{\end{subequations}}

\newcommand{\bmulti}{\begin{multline}}   %

\newcommand{\beqm}{\begin{mathletters}}   %

\newcommand{\eeqm}{\end{mathletters}}   %

\newcommand{\change}[1]{{#1}}

\begin{document}

\titlerunning{Is there a need to measure the cosmic microwave
  background temperature more accurately?}
\title{Is there a need and another way to measure the cosmic microwave
background temperature more accurately?}

\author{J. Chluba\inst{1} \and R.A. Sunyaev\inst{1,2}}
\authorrunning{Chluba \and Sunyaev}

\institute{Max-Planck-Institut f\"ur Astrophysik, Karl-Schwarzschild-Str. 1,
85741 Garching bei M\"unchen, Germany 
\and 
Space Research Institute, Russian Academy of Sciences, Profsoyuznaya 84/32,
117997 Moscow, Russia
}

\offprints{J. Chluba, 
\\ \email{jchluba@mpa-garching.mpg.de}
}

\date{Received 3 July 2007 / Accepted 4 December 2007}

\abstract{The recombination history of the Universe depends exponentially on
  the temperature, $T_0$, of the cosmic microwave background. Therefore tiny
  changes of $T_0$ are expected to lead to significant changes in the
  free electron fraction.  
  Here we show that even the current $1\sigma$-uncertainty in the
  value of $T_0$ \change{results in} more than half a percent ambiguity in the
  ionization history, and more than $0.1\%$ uncertainty in the $TT$
  and $EE$ power spectra at small angular scales.
  We discuss how the value of $T_0$ affects the highly redshifted cosmological
  hydrogen recombination spectrum and demonstrate that $T_0$ could, in
  principle, be measured by looking at the low frequency distortions of the
  cosmic microwave background spectrum. For this no absolute measurements are
  necessary, but sensitivities on the level of $\sim 30\,$nK are required to
  extract the {\it quasi-periodic frequency-dependent signal} with typical
  $\Delta \nu/\nu\sim 0.1$ coming from cosmological recombination.
We also briefly mention the possibility of obtaining additional
information on the specific entropy of the Universe, and other cosmological
parameters.  
}
\keywords{Cosmic Microwave Background -- Cosmology: theory} 

\maketitle

\section{Introduction}
\label{sec:Intro}
The recombination history of the Universe \citep{Zeldovich68, Peebles68,
SeagerRecfast1999, Seager2000} depends exponentially on the exact value of
$T_0$ \citep{Sunyaev1970}. Therefore one expects that tiny changes of $T_0$
lead to significant modifications of the ionization history.
The temperature of the cosmic microwave background (CMB) was measured with
tremendously high accuracy using the {\sc Cobe / Firas} instrument: 
$T_0=2.725\pm 0.001\,$K \citep{Fixsen2002} or $\Delta T/T\sim 0.04\%$,
\change{where $1\,$mK is the $1\sigma$-error}.
This value is based on the results obtained with the {\sc Cobe / Firas} and
  the quoted uncertainties from other methods, e.g. by measuring the CMB
  dipole anisotropy with the {\sc Cobe / Dmr} instrument: $T_0=2.725\pm
  0.020\,$K \citep{Kogut1996} or its slightly improved value $T_0=2.725\pm
  0.012\,$K \citep{Mather1999}, are much larger, yielding $\Delta T/T\sim
  0.7\%$ and $\Delta T/T\sim 0.4\%$ respectively. Also within the {\sc Cobe /
  Firas} instrument a systematic difference of $\Delta T\sim 5\,$mK arose due
  to readout current heating \citep{Mather1999}, which was only understood
  later \citep{Fixsen1996, Mather1999, Fixsen2002}. The main goal of the
  present paper is to demonstrate that even changes of the order of a few mK
  lead to percent-level corrections in the ionization history, and that in
  principle there may be another {\it indirect}, but potentially accurate way
  to measure the CMB monopole temperature \change{using the cosmological
  recombination spectrum coming from redshifts $z\gtrsim 800$}.

The impact of the uncertainty in the \change{CMB} monopole temperature on the
  CMB angular power spectra was addressed well before the era of precision
  cosmology \citep{Hu1995} with the conclusion that the corresponding
  theoretical error is below $\sim 1\%$.
  However, the great success in observations of the CMB temperature and
polarization anisotropies \citep{Hinshaw2006, Page2006} renders this level of
uncertainty insufficient for the analysis and interpretation of future
CMB data, which will become available after the launch of the {\sc Planck}
Surveyor\footnote{www.rssd.esa.int/Planck}, or with {\sc Cmbpol}.
As explained in \citet{Seljak2003}, the ultimate goal is to achieve $\sim
0.1\%$ accuracy for the theoretical prediction of the CMB power spectra.
In particular when discussing the imprints of different inflationary
models on the power spectra, or when obtaining estimates of the key
cosmological parameters that reach sub-percent accuracy, this precision
becomes necessary.

Currently, in this context the largest uncertainty is considered to be due to
our understanding of the epoch of cosmological recombination
\citep[e.g.][]{Seljak2003}.
This fact has recently motived several studies on high precision computations
of the cosmological hydrogen \citep{Dubrovich2005, Chluba2006, Kholu2006,
Jose2006, Chluba2007, Chluba2007b, Chluba2007c}, and helium \citep{Wong2006b,
SwitzerHirata2007I, SwitzerHirata2007III, SwitzerHirata2007II, Kholu2007,
Jose2007} recombination history.
All the discussed additional physical processes lead to $\gtrsim 0.1\%$ level
corrections of the ionization history, which also partially cancel each other.
The overall theoretical uncertainty in the CMB temperature and polarization
power spectra, in particular at large $l$, still exceeds the level of $0.1\%$.
Here we show that \change{the current $1\sigma$-error} in the value of
$T_0$ \change{also} yields more than $0.1\%$ uncertainty in the $TT$ and $EE$
power spectra.

During the epoch of cosmological hydrogen recombination, roughly 5 photons are
emitted per recombined hydrogen atom \citep{Chluba2006b}. As realized earlier
\citep[e.g. see][]{Zeldovich68, Peebles68, Dubrovich1975, Bernshtein1977,
Dubrovich1995, Burgin2003, Dubrovich2004b, Kholu2005}, the amount of these
photons and their spectral distribution also depends on the parameters of the
Universe.
Here we demonstrate that for the cosmological recombination spectrum the
dependence on the CMB \change{monopole} temperature, $T_0$, the total amount
of baryons, $\propto\Omega_{\rm b}h^2$, and the abundance ratio of helium to
hydrogen, $Y_{\rm p}$, are likely to be most important.
Basing our analysis on the code developed in \citet{Jose2006} and
\citet{Chluba2007}, we illustrate that in principle, by measuring the
cosmological recombination spectrum in the decimeter spectral band, one should
be able to determine the value of $T_0$.
Most importantly, for this {\it no absolute measurements} are necessary, but
sensitivities on the level of $30\,$nK are required to extract the
{\it quasi-periodic} signal with typical $\Delta \nu/\nu\sim 0.1$ coming from
cosmological recombination.
For observations of the CMB angular fluctuations, a sensitivity level of
$10\,$nK in principle can be achieved \citep{ReadheadPC}. However, here the
same signal is coming from every direction on the sky. Therefore one may use
\change{wide}-angle horns, \change{so that} one is dealing with a huge flux of
photons, carefully selecting regions on the sky that are cleanest with respect
to foreground signals.
%
 
\begin{figure}
\centering
\includegraphics[width=\columnwidth]
{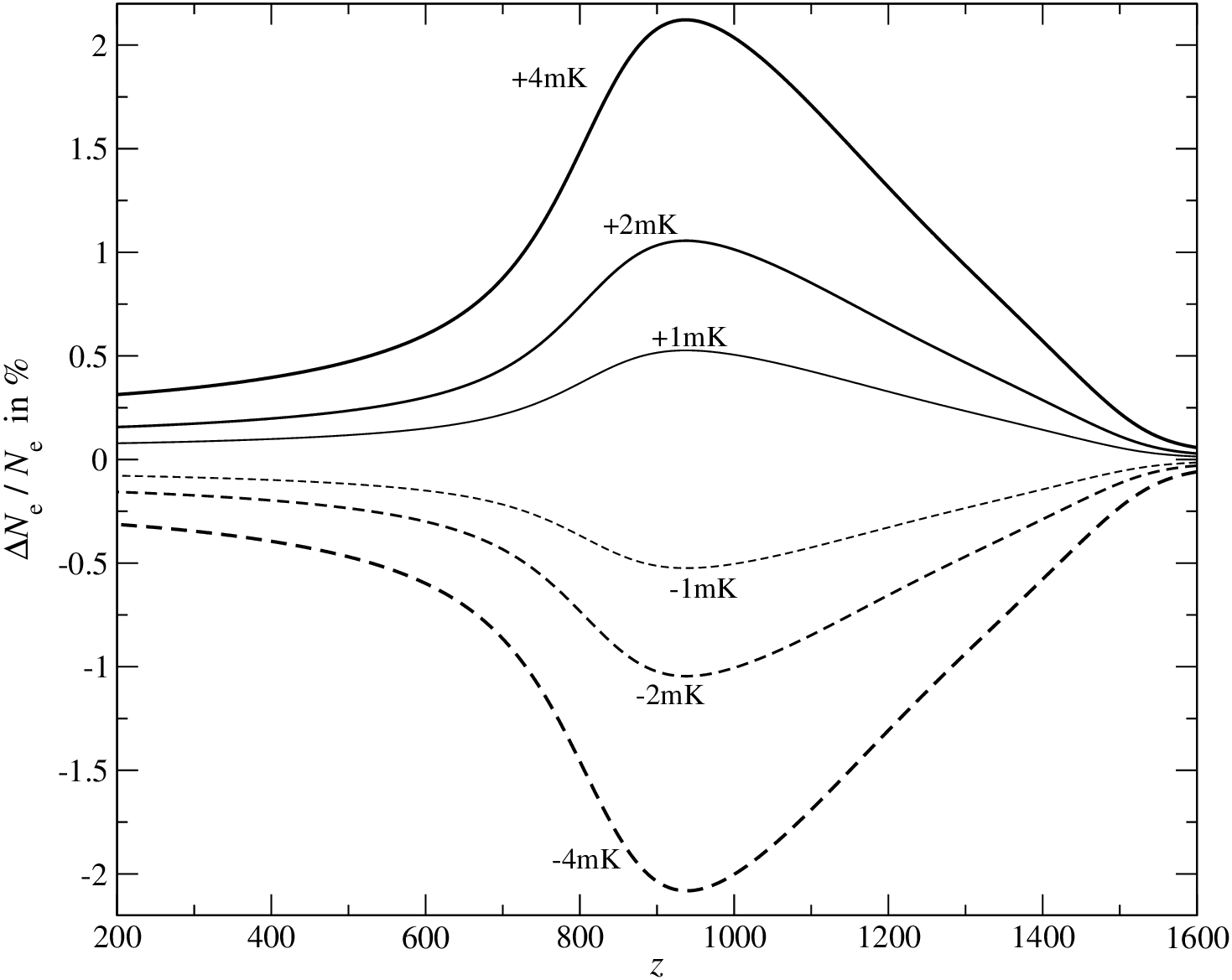}
\caption{Ambiguity in the ionization history due to uncertainty in the value
  of $T_0$. We changed $T_0=2.725\,$K by $\Delta T$ as labeled. All the other
  parameters were not altered.}
\label{fig:DXe}
\end{figure}
\begin{figure}
\centering
\includegraphics[width=\columnwidth]
{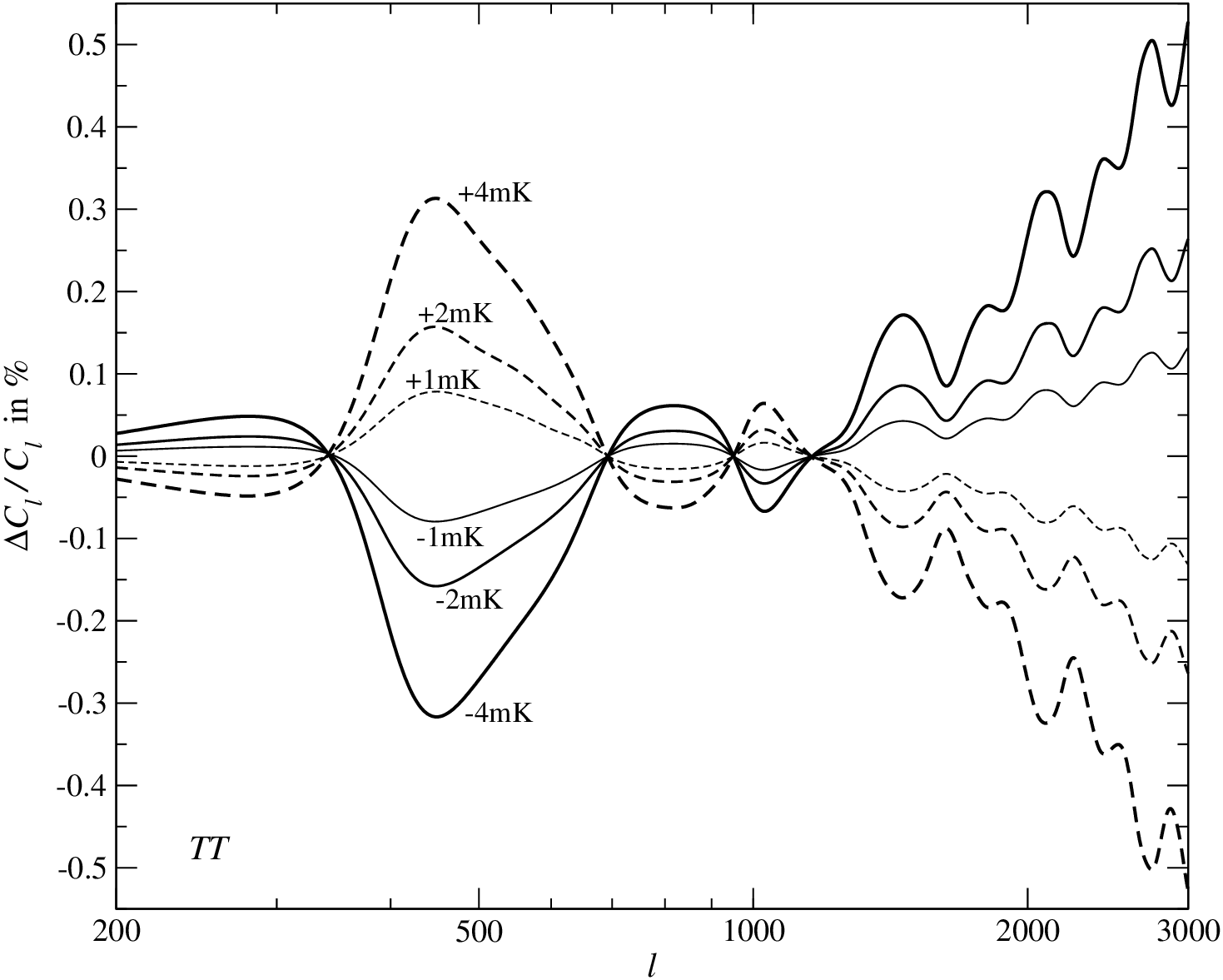}
\\
\includegraphics[width=\columnwidth]
{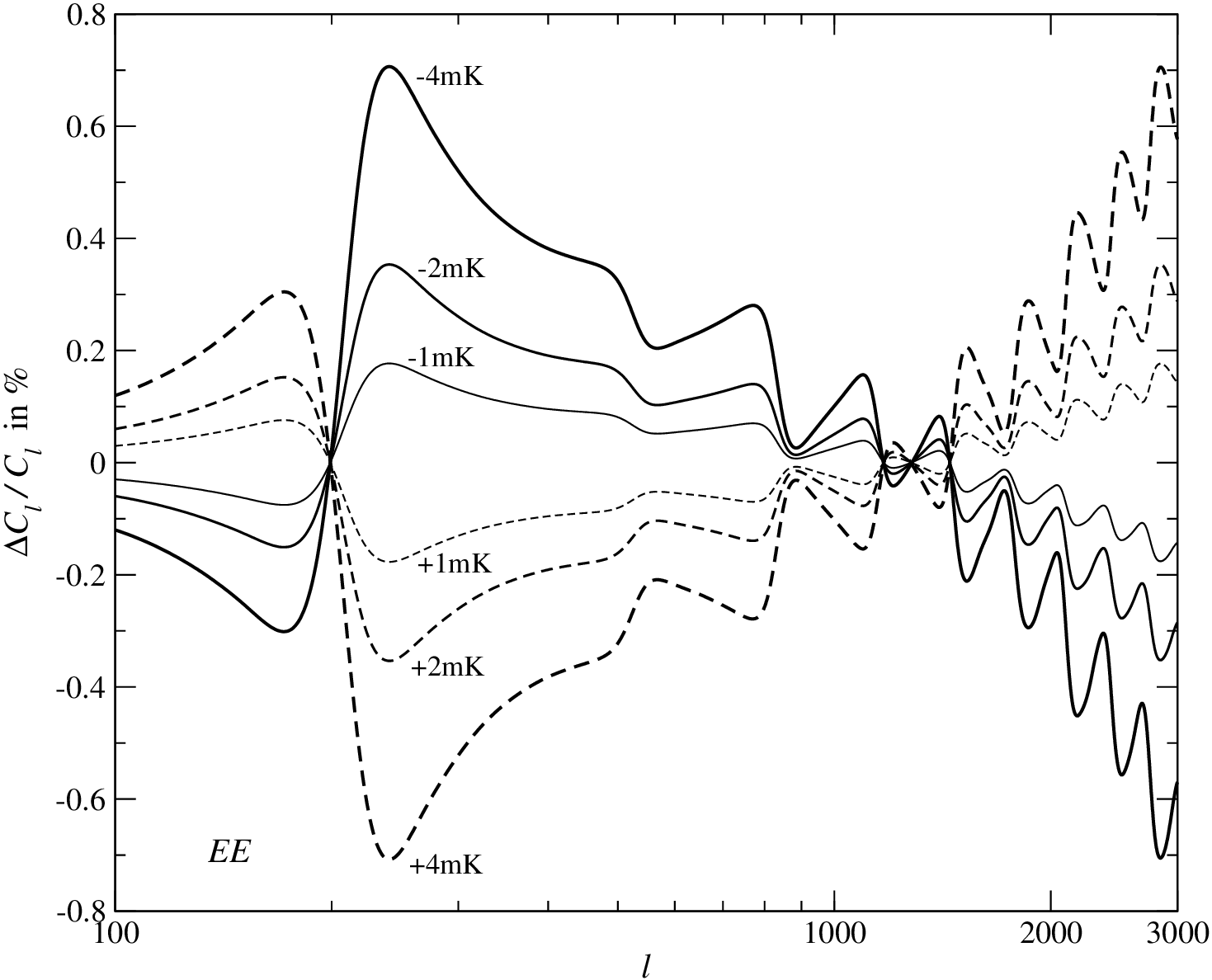}
\caption{Ambiguity in the temperature (top panel) and $E$-mode
  polarization (lower panel) power spectra due to uncertainty in the value of
  $T_0$. We changed $T_0=2.725\,$K by $\Delta T$ as labeled. All the other
  parameters were not modified.}
\label{fig:DCl}
\end{figure}
\section{Effect on the free electron fraction and CMB power spectra}
\label{sec:DXeDCl}
To study how the uncertainty in the value of $T_0$ affects the ionization
history and CMB temperature and polarization power spectra, we used the {\sc
  Cmbeasy} code \citep{Doran2005}. We ran our computations for a flat
$\Lambda$CDM cosmological concordance model 
%
\citep{WMAP_params}
with $T_0=2.725\,$K, $\Omega_{\rm cdm}=0.2234$, $\Omega_{\rm b}=0.0444$,
$h=0.71$ and $Y_{\rm p}=0.24$.

In Fig. \ref{fig:DXe} we show the uncertainty in the free electron fraction,
$N_{\rm e}$.
As expected, a slightly higher value of $T_0$ delays recombination, while a
smaller value shifts the time of recombination towards higher redshifts.
Even for $\Delta T=1\,$mK, the change in the ionization history exceeds half a
percent at $z\sim 950$. This shows that due to the exponential dependence of
$N_{\rm e}$ on $T_0$ \citep{Sunyaev1970} a tiny error of $\Delta T/T_0\sim
0.04\%$ is amplified by more than one order of magnitude.
The maximal change in the free electron fraction roughly scales as $\Delta
N_{\rm e}/N_{\rm e}\approx 15\,\Delta T/T_0\approx 0.55\% \times [\frac{\Delta
T}{1\text{mK}}]$ for $\Delta T/T_0 \ll 1$.

In Fig. \ref{fig:DCl} we show the corresponding uncertainty in the $TT$ and
$EE$ power spectra. At large scales, the positions of the first peaks are
slightly shifted, while at small scales ($l\gtrsim 1000$), the actual
modifications due to photon diffusion \citep{Silk1968} dominate. For earlier
recombination ($\Delta T < 0$), more power is left in the $TT$ power-spectrum
at small scales. On the other hand the amplitude of the $E$-mode polarization
is lower, likely due to less scattering \change{off} moving electrons.
In both cases the uncertainty is several times smaller than for the ionization
history: even for $\Delta T=\pm 4\,$mK, where $\Delta N_{\rm e}/N_{\rm e}$
exceeds 2\% close to its maximum, the power spectra are affected by less than
a percent at $l\leq 3000$.
For the $TT$ power spectrum one has $\Delta C_l/C_l\approx 2.1\,\Delta T/T_0$
at $l\sim 450$ and $\Delta C_l/C_l\approx -3.5\,\Delta T/T_0$ at $l\sim
3000$. Similarly for the $EE$ power spectrum one finds $\Delta C_l/C_l\approx
-4.8\,\Delta T/T_0$ at $l\sim 240$ and $\Delta C_l/C_l\approx 4.1 \,\Delta
T/T_0$ at $l\sim 3000$.

\change{
The relative change is growing towards larger $l$}, but at $l\gtrsim
1000-2000$ the signals connected with clusters of galaxies
(\citealt{Sunyaev1970, Sunyaev1972} and e.g. \citealt{Springel2001, Bjoern2006}), weak lensing,
foregrounds, and (especially at high frequencies) {\sc Scuba} sources
\citep{Haiman2000} and starforming, merging, dusty galaxies \citep{Righi2007}
will start to be more important.
Still, at $l\lesssim 1000-2000$ the multi-$l$ comparison of the observed CMB
power spectra with the predicted {\it theoretical mean model} over a {\it wide
range of angular scales} makes the discussed changes \change{potentially
important} once systematic and statistical errors decrease with the
improvements achieved using future CMB experiments.
Furthermore, the effect under discussion has a similar order of magnitude as
the corrections due to previously neglected physical processes (see
Introduction), so it should be included in the list of mechanisms that lead to
small modifications in the ionization history and CMB power spectra.
%
%

\begin{figure}
\centering
\includegraphics[width=\columnwidth]
{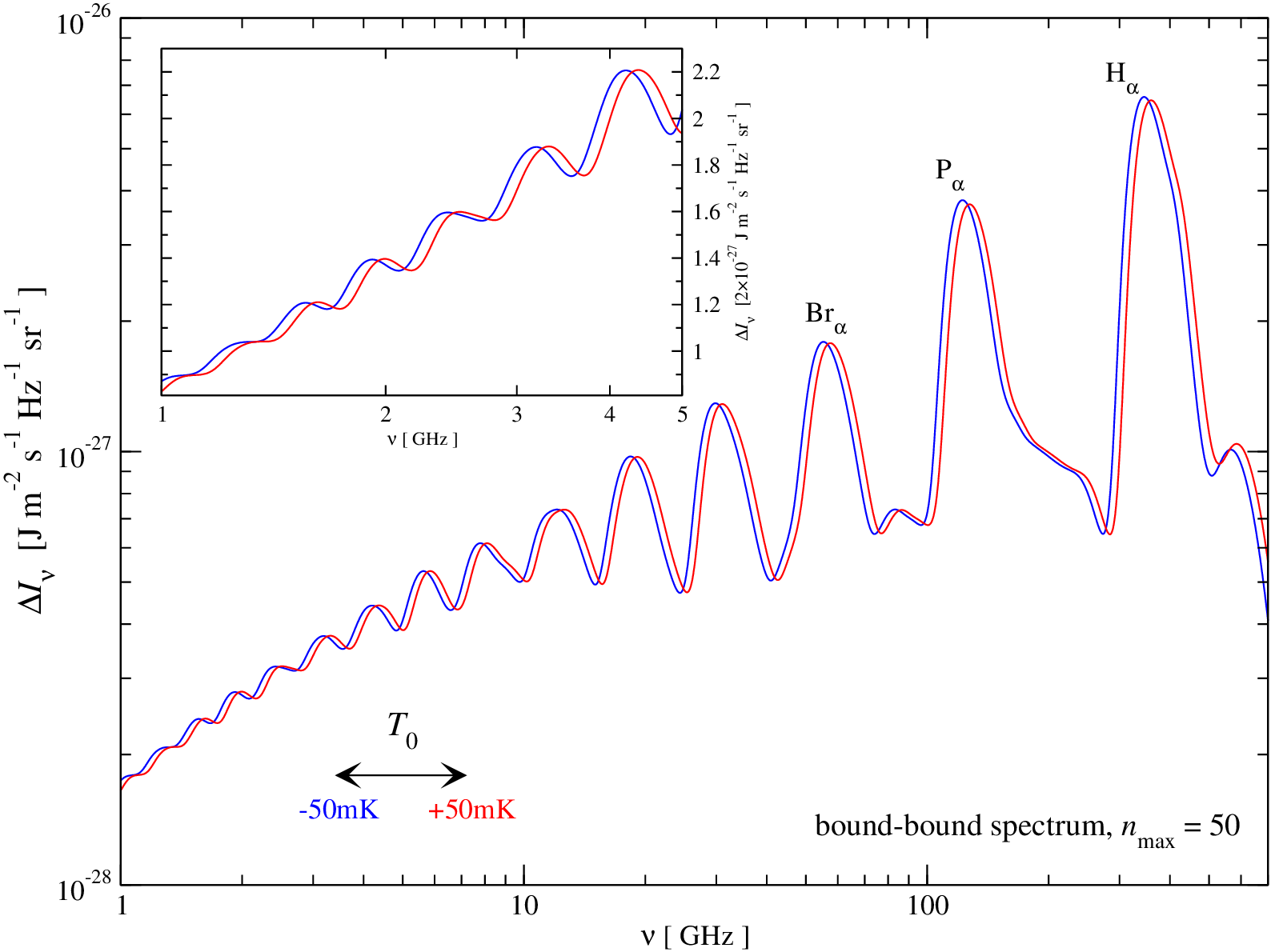}
\caption{Dependence of the bound-bound recombination spectrum on the value of
  $T_0$. We chose a large value for $\Delta T$ in order to illustrate the
  effect. All the other cosmological parameters remained unchanged. The
  results are based on computations including 50 shells for the hydrogen atom
  (see \citealt{Jose2006} and \citealt{Chluba2007} for computational details).
  The whole recombination spectrum is shifted along the frequency axis, but
  the overall normalization is not affected.}
\label{fig:DJ_T0}
\end{figure}
\begin{figure}
\centering
\includegraphics[width=\columnwidth]
{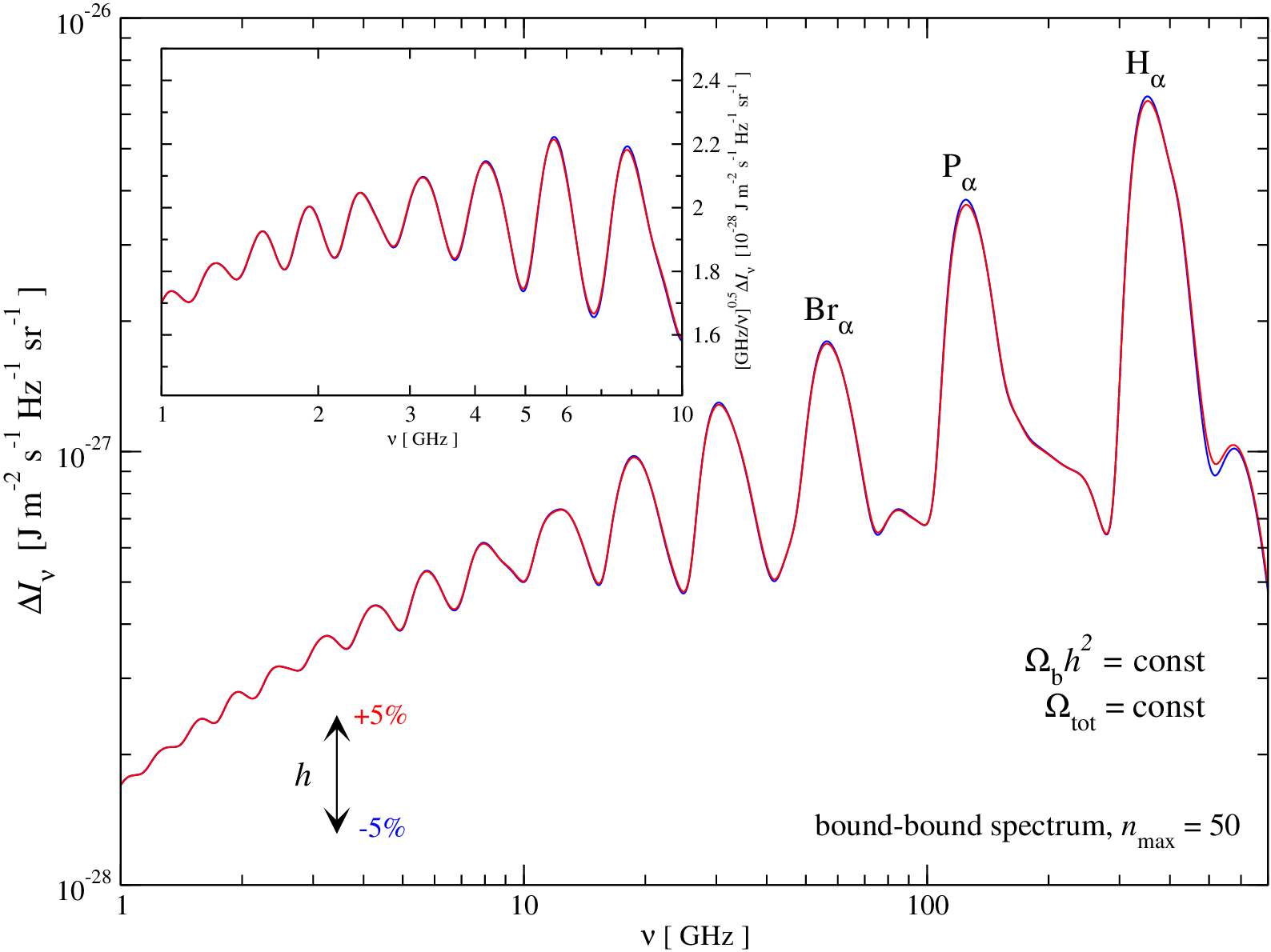}
\\
\includegraphics[width=\columnwidth]
{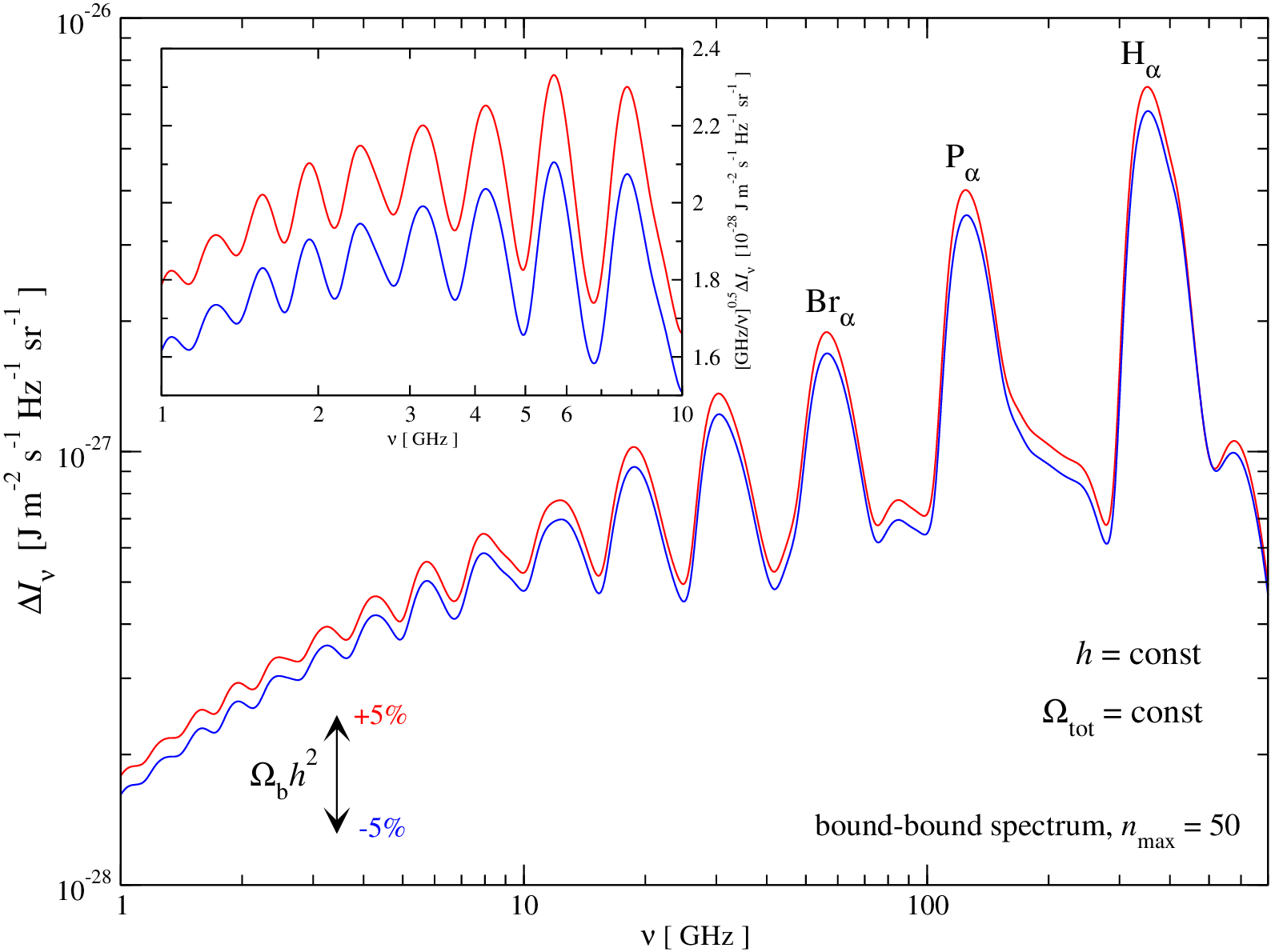}
\caption{Dependence of the bound-bound recombination spectrum on the number of
hydrogen nuclei, $N_{\rm p}\propto\Omega_{\rm b} h^2$. In the upper panel we
varied $h$, but fixed $\Omega_{\rm b} h^2$ and $\Omega_{\rm tot}=1$, while in
the lower panel we changed $\Omega_{\rm b}$ keeping $h$ and $\Omega_{\rm
tot}=1$ fixed.
All results are based on computations including 50 shells for the hydrogen
atom (see \citealt{Jose2006} and \citealt{Chluba2007} for computational
details).}
\label{fig:DJ_cosm}
\end{figure}
\section{Dependence of the recombination spectrum on different cosmological parameters}
\label{sec:DJ_cosm}
In this Sect. we {\it illustrate} the impact of different
cosmological parameters on the hydrogen recombination spectrum. We restrict
ourselves to the bound-bound emission spectrum and included 50 shells for the
hydrogen atom into our computations. A more rigorous investigation is in
preparation, however, the principle conclusions should not be affected.

In Fig.~\ref{fig:DJ_T0} we illustrate the dependence of the hydrogen
recombination spectrum on the value of $T_0$. 
The value of $T_0$ mainly defines the time of recombination, and consequently
when most of the emission in each transition appears. This leads to a
dependence of the line positions on $T_0$, but the total intensity in each
transition (especially at frequencies $\nu \lesssim 30\,$GHz) remains
practically the same. \change{The fractional shift of the low frequency
spectral features along the frequency axis scales roughly as $\Delta
\nu/\nu\sim\Delta T/T_0$.} Hence $\Delta T\sim1\,$mK implies
$\Delta\nu/\nu\sim 0.04\,\%$ or $\Delta\nu\sim 1\,$MHz at $2\,$GHz.
Since the maxima and minima of the line features due to the large duration of
recombination are rather broad ($\sim 10-20\%$), it is probably better to look
for these shifts close to the steep parts of the lines, where the derivatives
of the spectral distortion due to hydrogen recombination are largest.
It is also important to mention that the hydrogen recombination spectrum is
shifted as a {\it whole}, allowing to increase the significance of a
measurement by considering many spectral features at several frequencies.

The upper panel of Fig.~\ref{fig:DJ_cosm} shows that at low frequencies the
cosmological hydrogen recombination spectrum is practically independent of the
value of $h$. Only the features due to the Lyman, Balmer, Paschen and
Brackett series are slightly modified.
  This is connected to the fact that $h$ affects the ratio of the atomic time
  scales to the expansion time. Therefore changing $h$ affects the escape rate
  of photons in the Lyman-$\alpha$ transition and the relative importance of
  the 2s-1s transition. For transitions among highly excited states it is not
  crucial via which channel the electrons finally reach the ground state of
  hydrogen and hence the modifications of the recombination spectrum at low
  frequencies due to changes of $h$ are small.
  Changes of $\Omega_{\rm m}h^2$ should affect the recombination spectrum for
  the same reason.

  The lower panel in Fig.~\ref{fig:DJ_cosm} illustrates the dependence of the
  hydrogen recombination spectrum on $\Omega_{\rm b}h^2$. \change{The total
  number} of photons released during hydrogen recombination is directly
  related to the total number of hydrogen nuclei \citep[e.g.][]{Chluba2006b}.
  Therefore one expects that the overall normalization of the recombination
  spectrum depends on the total number of baryons, $N_{\rm
  b}\propto\Omega_{\rm b}h^2$, and the helium to hydrogen abundance ratio,
  $Y_{\rm p}$.
  Varying $\Omega_{\rm b}h^2$ indeed leads to a change in the overall
  amplitude $\propto \Delta(\Omega_{\rm b}h^2)/(\Omega_{\rm b}h^2)$.
  Similarly, changes of $Y_{\rm p}$ would affect the normalization of the
  hydrogen recombination spectrum, however it is important to {\it also}
  take the helium recombination spectrum into account.
%
%
Changing $Y_{\rm p}$ will affect the relative contribution of the helium to
  the hydrogen recombination spectrum. Since the physics of helium
  recombination is different than in the case of hydrogen (e.g. the spectrum
  of neutral helium is more complicated; helium recombination occurs at
  earlier times, when the medium was hotter;
  $\ion{He}{iii}\rightarrow\ion{He}{ii}$ is more rapid, so the recombination
  lines are narrower\footnote{Our computations show that the broadening due to
  electron scattering is $\Delta\nu/\nu \lesssim 7\%$, so the overall width of
  the $\ion{He}{iii}\rightarrow\ion{He}{ii}$-features is still smaller than
  the cosmological hydrogen recombination lines \citep{Jose2007}}), one can
  expect to find direct evidence of the presence of helium in the full
  recombination spectrum. These might be used to quantify the total amount of
  helium during the epoch of recombination, well before the first appearance
  of stars.

\section{Discussion and conclusions}
\label{sec:dis}
In Sect. \ref{sec:DXeDCl} we have shown that the \change{$1\sigma$-error} in
the exact value of $T_0$ leads to more than half a percent ambiguity in the
recombination history, resulting in more than $0.1\%$ uncertainty in the $TT$
and $EE$ power spectra. In the analysis of future CMB data one can easily
include this {\it additional} source of uncertainty by considering the value
of $T_0$ as a well-constrained free parameter.
%

It is clear that at the $\gtrsim 0.1\%$ level, the uncertainties in the helium
  abundance ratio, $Y_{\rm p}$, and the effective number of neutrinos,
  $N_\nu$, should also be considered \citep[e.g.][]{Hu1995,
  Olive1995,Cyburt2004,Steigman2006}. In particular the value of $Y_{\rm p}$
  has a strong impact, directly affecting the peak of the Thomson
  visibility function \citep{Sunyaev1970} due to the dependence of the number
  of hydrogen nuclei, $N_{\rm H}$, for a fixed number of baryons, on the
  helium abundance ratio ($N_{\rm H} \propto [1-Y_{\rm p}]$).
It is important to mention that the effects connected with the angular
  fluctuations of the CMB temperature ($\Delta T/T\sim \text{few}\times
  10^{-5}$), as observed with {\sc Cobe} and {\sc Wmap}, is expected to be one
  order of magnitude smaller.

In Sect. \ref{sec:DJ_cosm} we discussed the dependence of the cosmological
hydrogen recombination spectrum on different cosmological parameters. We have
illustrated that changes in the value of the CMB temperature, $T_0$, lead to
an overall shift of the hydrogen recombination spectrum along the frequency
axis, leaving the total normalization practically unchanged. On the other hand
changes of the total amount of baryons, $\propto\Omega_{\rm b}h^2$, mainly
affects the amplitude of the recombination signal.
This in principle should allow measurement of both parameters separately.
However, a more detailed analysis is required, also including the lines from
cosmological helium recombinations \change{\citep{Jose2007}}, and the
free-bound and two-photon contributions.
In this way it may be possible to construct an accurate {\it spectral
  template}, which can be used to disentangle the cosmological recombination
spectrum from other astrophysical sources. 

In addition to the interesting potential of constraining cosmological
parameters (without suffering from limitations set by cosmic
variance), an observation of the radiation from the epoch of recombination
will yield the final proof that our theoretical understanding of how the
Universe became neutral is correct.
This aspect is particularly important when considering the possibility of
non-standard recombination scenarios \citep{Peebles2000, Doroshkevic2003,
Bean2003, Bean2007}, which would also affect the results obtained here.

\acknowledgement{The authors wish to thank E.~E.~Kholupenko for his comments
    on the paper, and \change{J.~R.~Bond, C.~Hern\'{a}ndez-Monteagudo,
    J.~A.~Rubi\~no-Mart\'{\i}n and B.~D.~Wandelt} for useful discussions. We
    are grateful for discussions on experimental possibilities with
    J.~E.~Carlstrom, D.~J.~Fixsen, A.~Kogut, M.~Pospieszalski, A.~Readhead,
    E.~J.~Wollack and especially J.~C.~Mather.}

\bibliographystyle{aa} 
\bibliography{Lit}

\begin{thebibliography}{49}
\expandafter\ifx\csname natexlab\endcsname\relax\def\natexlab#1{#1}\fi

\bibitem[{{Bean} {et~al.}(2003){Bean}, {Melchiorri}, \& {Silk}}]{Bean2003}
{Bean}, R., {Melchiorri}, A., \& {Silk}, J. 2003, \prd, 68, 083501

\bibitem[{{Bean} {et~al.}(2007){Bean}, {Melchiorri}, \& {Silk}}]{Bean2007}
{Bean}, R., {Melchiorri}, A., \& {Silk}, J. 2007, \prd, 75, 063505

\bibitem[{{Bennett} {et~al.}(2003){Bennett}, {Halpern}, {Hinshaw}, {Jarosik},
  {Kogut}, {Limon}, {Meyer}, {Page}, {Spergel}, {Tucker}, {Wollack}, {Wright},
  {Barnes}, {Greason}, {Hill}, {Komatsu}, {Nolta}, {Odegard}, {Peiris}, \&
  {Verde}}]{WMAP_params}
{Bennett}, C.~L., {Halpern}, M., {Hinshaw}, G., {et~al.} 2003, \apjs, 148, 1

\bibitem[{{Bernshtein} {et~al.}(1977){Bernshtein}, {Bernshtein}, \&
  {Dubrovich}}]{Bernshtein1977}
{Bernshtein}, I.~N., {Bernshtein}, D.~N., \& {Dubrovich}, V.~K. 1977, Soviet
  Astronomy, 21, 409

\bibitem[{{Burgin}(2003)}]{Burgin2003}
{Burgin}, M.~S. 2003, Astronomy Reports, 47, 709

\bibitem[{{Chluba} {et~al.}(2007){Chluba}, {Rubi{\~n}o-Mart{\'{\i}}n}, \&
  {Sunyaev}}]{Chluba2007}
{Chluba}, J., {Rubi{\~n}o-Mart{\'{\i}}n}, J.~A., \& {Sunyaev}, R.~A. 2007,
  \mnras, 374, 1310

\bibitem[{{Chluba} \& {Sunyaev}(2006{\natexlab{a}})}]{Chluba2006b}
{Chluba}, J. \& {Sunyaev}, R.~A. 2006{\natexlab{a}}, \aap, 458, L29

\bibitem[{{Chluba} \& {Sunyaev}(2006{\natexlab{b}})}]{Chluba2006}
{Chluba}, J. \& {Sunyaev}, R.~A. 2006{\natexlab{b}}, \aap, 446, 39

\bibitem[{{Chluba} \& {Sunyaev}(2007{\natexlab{a}})}]{Chluba2007b}
{Chluba}, J. \& {Sunyaev}, R.~A. 2007{\natexlab{a}}, \aap, 475, 109

\bibitem[{{Chluba} \& {Sunyaev}(2007{\natexlab{b}})}]{Chluba2007c}
{Chluba}, J. \& {Sunyaev}, R.~A. 2007{\natexlab{b}}, ArXiv e-prints, 705

\bibitem[{{Cyburt}(2004)}]{Cyburt2004}
{Cyburt}, R.~H. 2004, \prd, 70, 023505

\bibitem[{{Doran}(2005)}]{Doran2005}
{Doran}, M. 2005, Journal of Cosmology and Astro-Particle Physics, 10, 11

\bibitem[{{Doroshkevich} {et~al.}(2003){Doroshkevich}, {Naselsky}, {Naselsky},
  \& {Novikov}}]{Doroshkevic2003}
{Doroshkevich}, A.~G., {Naselsky}, I.~P., {Naselsky}, P.~D., \& {Novikov},
  I.~D. 2003, \apj, 586, 709

\bibitem[{{Dubrovich}(1975)}]{Dubrovich1975}
{Dubrovich}, V.~K. 1975, Soviet Astronomy Letters, 1, 196

\bibitem[{{Dubrovich} \& {Grachev}(2004)}]{Dubrovich2004b}
{Dubrovich}, V.~K. \& {Grachev}, S.~I. 2004, Astronomy Letters, 30, 657

\bibitem[{{Dubrovich} \& {Grachev}(2005)}]{Dubrovich2005}
{Dubrovich}, V.~K. \& {Grachev}, S.~I. 2005, Astronomy Letters, 31, 359

\bibitem[{{Dubrovich} \& {Stolyarov}(1995)}]{Dubrovich1995}
{Dubrovich}, V.~K. \& {Stolyarov}, V.~A. 1995, \aap, 302, 635

\bibitem[{{Fixsen} {et~al.}(1996){Fixsen}, {Cheng}, {Gales}, {Mather},
  {Shafer}, \& {Wright}}]{Fixsen1996}
{Fixsen}, D.~J., {Cheng}, E.~S., {Gales}, J.~M., {et~al.} 1996, \apj, 473, 576

\bibitem[{{Fixsen} \& {Mather}(2002)}]{Fixsen2002}
{Fixsen}, D.~J. \& {Mather}, J.~C. 2002, \apj, 581, 817

\bibitem[{{Haiman} \& {Knox}(2000)}]{Haiman2000}
{Haiman}, Z. \& {Knox}, L. 2000, \apj, 530, 124

\bibitem[{{Hinshaw} {et~al.}(2006){Hinshaw}, {Nolta}, {Bennett}, {Bean},
  {Dore'}, {Greason}, {Halpern}, {Hill}, {Jarosik}, {Kogut}, {Komatsu},
  {Limon}, {Odegard}, {Meyer}, {Page}, {Peiris}, {Spergel}, {Tucker}, {Verde},
  \& {Weiland}}]{Hinshaw2006}
{Hinshaw}, G., {Nolta}, M.~R., {Bennett}, C.~L., {et~al.} 2006, ArXiv
  Astrophysics e-prints

\bibitem[{{Hirata} \& {Switzer}(2007)}]{SwitzerHirata2007II}
{Hirata}, C.~M. \& {Switzer}, E.~R. 2007, ArXiv Astrophysics e-prints

\bibitem[{{Hu} {et~al.}(1995){Hu}, {Scott}, {Sugiyama}, \& {White}}]{Hu1995}
{Hu}, W., {Scott}, D., {Sugiyama}, N., \& {White}, M. 1995, \prd, 52, 5498

\bibitem[{{Kholupenko} \& {Ivanchik}(2006)}]{Kholu2006}
{Kholupenko}, E.~E. \& {Ivanchik}, A.~V. 2006, Astronomy Letters, 32, 795

\bibitem[{Kholupenko {et~al.}(2005)Kholupenko, Ivanchik, \&
  Varshalovich}]{Kholu2005}
Kholupenko, E.~E., Ivanchik, A.~V., \& Varshalovich, D.~A. 2005, Gravitation
  and Cosmology, 11, 161

\bibitem[{{Kholupenko} {et~al.}(2007){Kholupenko}, {Ivanchik}, \&
  {Varshalovich}}]{Kholu2007}
{Kholupenko}, E.~E., {Ivanchik}, A.~V., \& {Varshalovich}, D.~A. 2007, \mnras,
  378, L39

\bibitem[{{Kogut} {et~al.}(1996){Kogut}, {Banday}, {Bennett}, {Gorski},
  {Hinshaw}, {Jackson}, {Keegstra}, {Lineweaver}, {Smoot}, {Tenorio}, \&
  {Wright}}]{Kogut1996}
{Kogut}, A., {Banday}, A.~J., {Bennett}, C.~L., {et~al.} 1996, \apj, 470, 653

\bibitem[{{Mather} {et~al.}(1999){Mather}, {Fixsen}, {Shafer}, {Mosier}, \&
  {Wilkinson}}]{Mather1999}
{Mather}, J.~C., {Fixsen}, D.~J., {Shafer}, R.~A., {Mosier}, C., \&
  {Wilkinson}, D.~T. 1999, \apj, 512, 511

\bibitem[{{Olive} \& {Steigman}(1995)}]{Olive1995}
{Olive}, K.~A. \& {Steigman}, G. 1995, \apjs, 97, 49

\bibitem[{{Page} {et~al.}(2006){Page}, {Hinshaw}, {Komatsu}, {Nolta},
  {Spergel}, {Bennett}, {Barnes}, {Bean}, {Dore'}, {Halpern}, {Hill},
  {Jarosik}, {Kogut}, {Limon}, {Meyer}, {Odegard}, {Peiris}, {Tucker}, {Verde},
  \& {Weiland}}]{Page2006}
{Page}, L., {Hinshaw}, G., {Komatsu}, E., {et~al.} 2006, ArXiv Astrophysics
  e-prints

\bibitem[{{Peebles}(1968)}]{Peebles68}
{Peebles}, P.~J.~E. 1968, \apj, 153, 1

\bibitem[{{Peebles} {et~al.}(2000){Peebles}, {Seager}, \& {Hu}}]{Peebles2000}
{Peebles}, P.~J.~E., {Seager}, S., \& {Hu}, W. 2000, \apjl, 539, L1

\bibitem[{{Readhead}(2007)}]{ReadheadPC}
{Readhead}, A. 2007, private communication

\bibitem[{{Righi} {et~al.}(2007){Righi}, {Hernandez-Monteagudo}, \&
  {Sunyaev}}]{Righi2007}
{Righi}, M., {Hernandez-Monteagudo}, C., \& {Sunyaev}, R. 2007, ArXiv e-prints,
  707

\bibitem[{{Rubi{\~n}o-Mart{\'{\i}}n} {et~al.}(2006){Rubi{\~n}o-Mart{\'{\i}}n},
  {Chluba}, \& {Sunyaev}}]{Jose2006}
{Rubi{\~n}o-Mart{\'{\i}}n}, J.~A., {Chluba}, J., \& {Sunyaev}, R.~A. 2006,
  \mnras, 371, 1939

\bibitem[{{Rubino-Martin} {et~al.}(2007){Rubino-Martin}, {Chluba}, \&
  {Sunyaev}}]{Jose2007}
{Rubino-Martin}, J.~A., {Chluba}, J., \& {Sunyaev}, R.~A. 2007, ArXiv e-prints,
  711

\bibitem[{{Sch{\"a}fer} {et~al.}(2006){Sch{\"a}fer}, {Pfrommer}, {Bartelmann},
  {Springel}, \& {Hernquist}}]{Bjoern2006}
{Sch{\"a}fer}, B.~M., {Pfrommer}, C., {Bartelmann}, M., {Springel}, V., \&
  {Hernquist}, L. 2006, \mnras, 370, 1309

\bibitem[{{Seager} {et~al.}(1999){Seager}, {Sasselov}, \&
  {Scott}}]{SeagerRecfast1999}
{Seager}, S., {Sasselov}, D.~D., \& {Scott}, D. 1999, \apjl, 523, L1

\bibitem[{{Seager} {et~al.}(2000){Seager}, {Sasselov}, \& {Scott}}]{Seager2000}
{Seager}, S., {Sasselov}, D.~D., \& {Scott}, D. 2000, \apjs, 128, 407

\bibitem[{{Seljak} {et~al.}(2003){Seljak}, {Sugiyama}, {White}, \&
  {Zaldarriaga}}]{Seljak2003}
{Seljak}, U., {Sugiyama}, N., {White}, M., \& {Zaldarriaga}, M. 2003, \prd, 68,
  083507

\bibitem[{{Silk}(1968)}]{Silk1968}
{Silk}, J. 1968, \apj, 151, 459

\bibitem[{{Springel} {et~al.}(2001){Springel}, {White}, \&
  {Hernquist}}]{Springel2001}
{Springel}, V., {White}, M., \& {Hernquist}, L. 2001, \apj, 549, 681

\bibitem[{{Steigman}(2006)}]{Steigman2006}
{Steigman}, G. 2006, Journal of Cosmology and Astro-Particle Physics, 10, 16

\bibitem[{{Sunyaev} \& {Zeldovich}(1970)}]{Sunyaev1970}
{Sunyaev}, R.~A. \& {Zeldovich}, Y.~B. 1970, Astrophysics and Space Science, 7,
  3

\bibitem[{{Sunyaev} \& {Zeldovich}(1972)}]{Sunyaev1972}
{Sunyaev}, R.~A. \& {Zeldovich}, Y.~B. 1972, Comments on Astrophysics and Space
  Physics, 4, 173

\bibitem[{{Switzer} \& {Hirata}(2007{\natexlab{a}})}]{SwitzerHirata2007I}
{Switzer}, E.~R. \& {Hirata}, C.~M. 2007{\natexlab{a}}, ArXiv Astrophysics
  e-prints

\bibitem[{{Switzer} \& {Hirata}(2007{\natexlab{b}})}]{SwitzerHirata2007III}
{Switzer}, E.~R. \& {Hirata}, C.~M. 2007{\natexlab{b}}, ArXiv Astrophysics
  e-prints

\bibitem[{{Wong} \& {Scott}(2007)}]{Wong2006b}
{Wong}, W.~Y. \& {Scott}, D. 2007, \mnras, 375, 1441

\bibitem[{{Zeldovich} {et~al.}(1968){Zeldovich}, {Kurt}, \&
  {Syunyaev}}]{Zeldovich68}
{Zeldovich}, Y.~B., {Kurt}, V.~G., \& {Syunyaev}, R.~A. 1968, Zhurnal
  Eksperimental noi i Teoreticheskoi Fiziki, 55, 278

\end{thebibliography}

\end{document}